\documentclass[preprint,showpacs,preprintnumbers,amsmath,amssymb]{revtex4}
\usepackage{graphicx}
\usepackage{dcolumn}
\usepackage{bm}
\usepackage{CJK}
\usepackage{amsfonts}
\usepackage{psfrag}
\usepackage{wrapfig}
\usepackage{subfigure}
\usepackage{makeidx}
\usepackage{multirow}
\usepackage{epsf}
\usepackage{hyperref}
\usepackage{amsmath}

\usepackage{cases}

\begin{document}

\newcommand{\eq}[1]{Eq.(\ref{#1})}
\newcommand{\ud}{\,\mathrm{d}\,}
\newcommand{\rmnum}[1]{\romannumeral #1}
\newcommand{\Rmnum}[1]{\uppercase\expandafter{\romannumeral #1}}

\title{Spontaneous emission of a two-level static atom coupling with the electromagnetic vacuum fluctuations outside a high-dimensional Einstein Gauss-Bonnet black hole}

\author{Ming Zhang$^{1}$, Zhan-Ying Yang$^{1}$, Rui-Hong Yue$^{2}$\footnote{yueruihong@nbu.edu.cn}}

\address{$^{1}$Department of Physics, Northwest University, Xi'an, 710069, China\\
$^{2}$Faculty of Science, Ningbo University, Ningbo 315211, China}

\begin{abstract}
  In present paper, by using the generalized DDC formalism, we investigate the spontaneous excitation of an static atom interacting with electromagnetic vacuum fluctuations outside a EGB black hole in $d$-dimensions. It shows that spontaneous excitation does not occur in Boulware vacuum, while exists in Unruh vacuum and Hartle-Hawking vacuum. As to the total rate of change of the atomic energy, it does not receive the contribution from the coupling constant of Gauss-Bonnet term at spatial infinity, only the dimensional parameter has the contribution to it. Near the event horizon, the coupling constant and the dimensional parameter both have contribution to the total rate of change of the atomic energy in all three kinds of vacuum.  We discuss the contribution of the coupling constant and dimensional factor to the results in three different kinds of spacetime in the last.
\end{abstract}


\maketitle

\section{INTRODUCTION}

Spontaneous emission is one of the most prominent effects in the interaction of atoms with radiation, which can be attributed to vacuum fluctuations, or radiation reaction, or a combination of them (\cite{Welton:1948zz}-\cite{Milonni:1975hc}). This ambiguous conception means that the contributions of vacuum fluctuations and radiation reaction can be chosen arbitrarily in a large extent, depending on the ordering of commuting atom and field variables. Dalibard, Dupont-Roc, and Cohen-Tannoudji (DDC) suggested that there exists a preferred operator ordering (\cite{Dalibard:1982kc}, \cite{Dalibard:1984lc}), only if one chooses a symmetric ordering are the distinct contributions of vacuum fluctuations and radiation reaction to the rate of change of an atomic observable separately Hermitian and able to possess an independent physical meaning. By using DDC formalism, Yu and Zhou studied the spontaneous excitation of a static multilevel atom coupled with electromagnetic vacuum fluctuations in four-dimensional Schwarzschild spacetime and calculate the spontaneous emission rate (\cite{Zhou:2012eb}).

However, following the advent of string theory, extra dimensions were promoted from an interesting curiosity to a theoretical necessity since superstring theory requires an eleven-dimensional spacetime to be consistent from a quantum point of view (\cite{Horava:1996ma}-\cite{Randall:1999vf}). Among the higher curvature gravities, the most extensively studied theory is the so-called Gauss-Bonnet gravity (\cite{Lanczos:1938sf} -\cite{Wheeler:1985qd}), which naturally emerges when we want to generalize Einstein's theory in higher dimensions by keeping all characteristics of usual general relativity excepting the linear dependence of the Riemann tensor. Therefore, we are curious to know what happens if the static multilevel atom coupled with electromagnetic vacuum fluctuations in high-dimensional Einstein Gauss-Bonnet black hole.

In this paper, we apply this generalized DDC formalism to investigate the spontaneous excitation of an static atom interacting with electromagnetic vacuum fluctuations outside a EGB black hole in $d$-dimensions and calculate the total rate of change of the mean energy. We discuss the stability of a two-level static atom in ground state and excited state of $d$-dimensional EGB black hole in Boulware vacuum, Unruh vacuum and Hartle-Hawking vacuum. We also analyse the contribution of the coupling constant and dimensional factor of the EGB gravity to the total rate of change when the coupling constant takes three different limit which represent three kinds of different spacetime.

\section{SETUP}

We consider a two-level atom in interaction with vacuum electromagnetic fluctuations outside a high-dimensional Einstein-Gauss-Bonnet black hole.
The action of EGB gravity can be written as
\begin{equation}
  S=\int\ud^{p+2}x~\sqrt{-g}(R+\alpha \mathcal {L}_2)
\end{equation}
where the coupling constant $\alpha$ can be regarded as the inverse of string tension and be assumed $\alpha>0$ in this paper.
The Gauss-Bonnet term is given by
\begin{equation}
  \mathcal {L}_2=R^2-4R_{ab}R^{ab}+R_{abcd}R^{abcd}
\end{equation}

The metric for the exterior region of the $d$-dimensional Einstein-Gauss-Bonnet gravity $(d=p+2)$ is given by:
\begin{equation}
  \ud s^2=f(r)\ud t^2-f(r)^{-1}\ud r^2-r^2\ud s_p^2
\end{equation}
where $ds_p^2$ is the line element of a unit $p$-sphere. The function $f(r)$ in the $d$-dimensional EGB gravity is
\begin{equation}
  f(r)=1+\frac{r^2}{2\alpha}(\sqrt{1-\frac{8\alpha M}{r^{d-1}}}-1)
\end{equation}
with $M$ being the mass of the black hole. We assume a pointlike two-level atom on the trajectory, which guarantees the existence of stationary atomic states. The Hamiltonian that describes the time evolution of the atom-field interacting system can be written as
\begin{equation}
  H(\tau)=H_A(\tau)+H_F(\tau)+H_I(\tau)
\end{equation}

Here $H_A(\tau)$ is the atom's undisturbed Hamiltonian, and it is given by (\cite{Dicke:1954zz})
\begin{equation}
  H_A(\tau)=\sum_n \omega_n \sigma_{nn}(\tau)
\end{equation}
where $\omega_n$ represents energy, and $\sigma_{nn}=|n\rangle\langle n|$.
$H_F(\tau)$ is the hamiltonian that decides the evolution of the free quantum electromagnetic field
\begin{equation}
  H_F(\tau)=\sum_{\vec{k}}\omega_{\vec{k}}a_{\vec{k}}^{\dag}a_{\vec{k}}\frac{\ud t}{\ud \tau}
\end{equation}
where $\vec{k}$ denotes the wave vector and polarization of the field modes.
We couple the two-level atom and the quantum electromagnetic field in the multipolar coupling scheme (\cite{Audretsch:1994yz}, \cite{Compagno:1995pz})
\begin{equation}
  H_I(\tau)=-e\textbf{r}(\tau)\cdot \textbf{E}(x(\tau))=-e\sum_{mn}\textbf{r}_{mn}\cdot \textbf{E}(x(\tau))\sigma_{mn}(\tau)
\end{equation}
where $e$ is the electron electric charge, $e\textbf{r}$ the atomic electric dipole moment, $x(\tau)$ the spacetime coordinates of the atom.

The equation of motion in the interaction representation for an arbitrary atomic observable $O(\tau)$, using symmetric ordering which was proposed by DDC, can be split into the vacuum fluctuations and the radiation reaction contributions,
\begin{equation}
  \frac{\ud O(\tau)}{\ud \tau}=(\frac{\ud O(\tau)}{\ud \tau})_{vf}+(\frac{\ud O(\tau)}{\ud \tau})_{rr}
\end{equation}

Now we assume that the initial state is a vacuum state $|0\rangle$, while the atom is in the state $|b\rangle$. We also assume the atom is polarized along the radial direction defined by the position of the atom relative to the black hole spacetime rotational killing fields. Then we only need to calculate the contributions in the $r$-direction. Our main aim is to identify the contributions of vacuum fluctuations and radiation reaction in the evolution of the atom's excitation energy, which is given by the expectation value of $H_A$ (\cite{Dalibard:1982kc}, \cite{Dalibard:1984lc})
\begin{eqnarray}
  \langle \frac{\ud H_A(\tau)}{\ud \tau} \rangle_{vf}=2ie^2\int_{\tau_0}^{\tau}\ud \tau' C^F(x(\tau),x(\tau'))\frac{\ud}{\ud \tau}\chi_b^A(\tau,\tau'), \label{equ:vf}\\
  \langle \frac{\ud H_A(\tau)}{\ud \tau} \rangle_{rr}=2ie^2\int_{\tau_0}^{\tau}\ud \tau' \chi^F(x(\tau),x(\tau'))\frac{\ud}{\ud \tau}C_b^A(\tau,\tau'), \label{equ:rrr}
\end{eqnarray}
where the statistical functions of the electromagnetic field $C^F$ and $\chi^F$ are also called symmetric correlation function and linear susceptibility of the field. the radial components of these function are defined as
\begin{eqnarray}
  &&C^F(x(\tau),x(\tau'))=\frac{1}{2}\langle 0|\{\textbf{E}_r^f(x(\tau)),\textbf{E}_r^f(x(\tau'))\}|0\rangle, \label{equ:cf}\\
  &&\chi^F(x(\tau),x(\tau'))=\frac{1}{2}\langle 0|[\textbf{E}_r^f(x(\tau)),\textbf{E}_r^f(x(\tau'))]|0\rangle \label{equ:xf}
\end{eqnarray}
We will give the exact form of the radial components $\textbf{E}_r^f(x(\tau))$ in the next section. $\{,\}$ denotes the anti-commutator.

Analogously, the symmetric correlation function and the linear susceptibility of the atom $C_b^A(\tau,\tau')$ and $\chi_b^A(\tau,\tau')$, are defined as
\begin{eqnarray}
  &&C_b^A(\tau,\tau')=\frac{1}{2}\langle b|\{\textbf{r}^f(\tau),\textbf{r}^f(\tau')\}|b\rangle, \\
  &&\chi_b^A(\tau,\tau')=\frac{1}{2}\langle b|[\textbf{r}^f(\tau),\textbf{r}^f(\tau')]|b\rangle
\end{eqnarray}
which are characterized by the atom itself. Explicitly, the statistical functions of the atom can be given as
\begin{eqnarray}
  &&C_b^A(\tau,\tau')=\frac{1}{2}\sum_d |\langle b|\textbf{r}(0)|d\rangle|^2[e^{i\omega_{bd}(\tau-\tau')}+e^{-i\omega_{bd}(\tau-\tau')}], \label{equ:cab}\\
  &&\chi_b^A(\tau,\tau')=\frac{1}{2}\sum_d |\langle b|\textbf{r}(0)|d\rangle|^2[e^{i\omega_{bd}(\tau-\tau')}-e^{-i\omega_{bd}(\tau-\tau')}] \label{equ:xab}
\end{eqnarray}
where $\omega_{bd}=\omega_b-\omega_d$ and the sum extends over a complete set of atomic states.

Now we can clear see that the quantization of electromagnetic fields outside a $d$-dimensional EGB black hole and the specification of vacuum states are necessary to calculate the rate of the mean atomic energy.

In the previous work, we examined free quantum electrodynamics in static spherically symmetric spacetime of arbitrary dimensions in a modified Feynman gauge. We gave all of the physical modes functions which consummate the results in Ref\cite{Zhang:2013hqa}. However, in the computation of the function $ _B\langle0|\hat{E}_r(t)\hat{E}_r(t')|0\rangle_B$ with $\hat{E}_r=\hat{A}_{t;r}-\hat{A}_{r;t}$, we can neglect all other modes except physical modes \Rmnum{1} according to their expression. Then we only need to consider the contribution of physical modes \Rmnum{1} (\cite{Crispino:2000jx}). In Boulware vacuum state $|0\rangle_B$, the correlation function is
\begin{eqnarray}
  \label{equ:ee}
  _B\langle 0|\hat{E}_r(t)\hat{E}_r(t')|0\rangle_B&&=\frac{1}{4\pi}\sum_{l}\int_0^{\infty}\ud \omega ~\omega e^{-i\omega(t-t')} Y_{lm}Y_{lm}^* \nonumber\\
  &&\times [\overrightarrow{R}_l(\omega|r)\overrightarrow{R}_l^{\ast}(\omega|r')+\overleftarrow{R}_l(\omega|r)\overleftarrow{R}_l^{\ast}(\omega|r')]
\end{eqnarray}
Here $Y_{lm}$ is a scalar spherical harmonic function on the unit $p$-sphere satisfying
\begin{eqnarray}
  &&\nabla^2 Y_{lm}=-l(l+p-1)Y_{lm} , \\
  &&\sum_{m=-1}^l |Y_{lm}|^2=\frac{G(l)}{\Omega_{p}}
\end{eqnarray}
where $\Omega_p$ is the volume of $S^{p}$ and $G(l)$ is the degeneracy of the eigenvalue $-l(l+p-2)$ of the Laplacian $\tilde{\Delta}$, which is given by
\begin{equation}
  G(l)=\frac{(2l+p-1)(l+p-2)!}{l!(p-1)!}.
\end{equation}

For the Unruh vacuum state $|0\rangle_U$ and the Hartle-Hawking vacuum state $|0\rangle_H$, the correlation functions are given respectively by
\begin{eqnarray}
  \label{equ:ee1}
  _U\langle 0|\hat{E}_r(t)\hat{E}_r(t')|0\rangle_U&&=\frac{1}{4\pi}\sum_{l}\int_0^{\infty}\ud \omega ~\omega e^{-i\omega(t-t')} Y_{lm}Y_{lm}^* \nonumber\\
  &&\times \Big[\frac{\overrightarrow{R}_l(\omega|r)\overrightarrow{R}_l^{\ast}(\omega|r')}{1-e^{-2\pi\omega/\kappa}}
  +\theta(\omega)\overleftarrow{R}_l(\omega|r)\overleftarrow{R}_l^{\ast}(\omega|r')\Big]
\end{eqnarray}
and
\begin{eqnarray}
  \label{equ:ee2}
  _H\langle 0|\hat{E}_r(t)\hat{E}_r(t')|0\rangle_H&&=\frac{1}{4\pi}\sum_{l}\int_0^{\infty}\ud \omega ~\omega \Big[e^{-i\omega(t-t')} Y_{lm}Y_{lm}^* \frac{\overrightarrow{R}_l(\omega|r)\overrightarrow{R}_l^{\ast}(\omega|r')}{1-e^{-2\pi\omega/\kappa}} \nonumber\\
   &&+e^{i\omega(t-t')} Y_{lm}^*Y_{lm}\frac{\overleftarrow{R}_l^{\ast}(\omega|r)\overleftarrow{R}_l(\omega|r')}{e^{2\pi\omega/\kappa}-1} \Big]
\end{eqnarray}
where $\kappa=f'(r_h)/2$ is the surface gravity of the black hole.

The radial function $R_l(\omega|r)$ satisfies the equation
\begin{equation}
  \label{equ:rr}
  [\frac{\omega^2}{f}-\frac{l(l+p-1)}{r^2}]R_{\omega l}^{(1n)}(r)+\frac{1}{r^2}\frac{\ud}{\ud r}[\sqrt{\frac{f}{h}} r^{2-p}\frac{\ud}{\ud r}(\frac{r^p}{\sqrt{fh}}R_{\omega l}^{(1n)}(r))]=0
\end{equation}
where the arrow represents modes incoming from the past null infinity $(\leftarrow)$ and coming out from the past horizon $(\rightarrow)$ .

Although it is difficult to obtain the exact solution of \eq{equ:rr}, we can give the behavior of the summation concerning the radial functions in the two asymptotic regions ($r\rightarrow r_h,~ r\rightarrow\infty$),
\begin{subequations}
  \begin{numcases}{\sum_l G(l)\cdot |\overleftarrow{R_l}|^2 \approx}
    \sum_l\frac{l(l+p-1)G(l)}{\omega^2 r_h^{p+2}}|\mathcal{T}_l(\omega)|^2, ~~r\rightarrow r_h  \label{equ:23a}\\
    \frac{(-2)^{2-p}\pi}{(\Gamma(\frac{p+1}{2}))^2}\cdot \frac{p}{p+1} \cdot \frac{\omega^{p}}{(\sqrt{g_{00}})^{p+2}}, ~~r\rightarrow \infty \label{equ:23b}
  \end{numcases}
  \begin{numcases}{\sum_l G(l)\cdot |\overrightarrow{R_l}|^2 \approx}
    \frac{\Gamma(\frac{p+2}{2}+\frac{2ir_h\omega}{a_1})\Gamma(\frac{p+2}{2}-\frac{2ir_h\omega}{a_1})\big(\Gamma(\frac{p+2}{2})\big)^2}{\xi^{p+2}\omega^2r^{p+2}\Gamma(p+2)(p-2)!\Gamma(\frac{2ir_h\omega}{a_1})\Gamma(-\frac{2ir_h\omega}{a_1})} ,~r\rightarrow r_h  \label{equ:23c}\\
    \sum_l\frac{l(l+p-1)G(l)}{\omega^2 r^{p+2}}|\mathcal{T}_l(\omega)|^2,~~r\rightarrow \infty \label{equ:23d}
  \end{numcases}
\end{subequations}
where $\mathcal{T}_l(\omega)$ represents the transmission coefficients and
\begin{eqnarray}
  &&\xi^2=\frac{r}{r_h}-1 \\
  &&a_1=-2+(p+1)\frac{r_h^2-\alpha}{r_h^2-2\alpha}
\end{eqnarray}

\section{SPONTANEOUS EXCITATION OF AN ATOM INTERACTING WITH THE VACUUM ELECTROMAGNETIC FIELD IN EGB GRAVITY}

\subsection{Boulware vacuum}
Put \eq{equ:ee} to \eq{equ:cf} and \eq{equ:xf}, we can obtain the symmetric correlation function and the linear susceptibility function
\begin{eqnarray}
  &&C^F(x(\tau),x(\tau'))=\frac{1}{8\pi\Omega_p}\int_0^{\infty}\ud\omega ~\omega(e^{-\frac{i\omega\Delta\tau}{\sqrt{g_{00}}}}+e^{\frac{i\omega\Delta\tau}{\sqrt{g_{00}}}})\sum_l G(l)(|\overrightarrow{R}|^2+|\overleftarrow{R}|^2), \label{equ:ncf}\\
  &&\chi^F(x(\tau),x(\tau'))=\frac{1}{8\pi\Omega_p}\int_0^{\infty}\ud\omega ~\omega(e^{-\frac{i\omega\Delta\tau}{\sqrt{g_{00}}}}-e^{\frac{i\omega\Delta\tau}{\sqrt{g_{00}}}})\sum_l G(l)(|\overrightarrow{R}|^2+|\overleftarrow{R}|^2) \label{equ:nxf}
\end{eqnarray}
Comparing the two functions with the one in the case of Schwarzschild black hole\cite{Zhou:2012eb}, we can see that the different between them is the term $\sum_l(2l+1)$ is replaced by the term $\sum_l G(l)$ since we study a high dimensional black hole and the coupling constant $\alpha$ is included in $g_{00}$. Following the same procedure\cite{Zhou:2012eb}, we can identify the contributions of vacuum fluctuations and radiation reaction in the evolution of the atom's excitation energy, then we can calculate the total rate of change of the atomic energy by adding them together
\begin{eqnarray}
  \Big\langle \frac{\ud H_A(\tau)}{\ud \tau} \Big\rangle_{tot}=\frac{-e^2g_{00}}{2\Omega_p}\sum_{\omega_b>\omega_d}|\langle b|\textbf{r}(0)|d\rangle|^2\omega_{bd}^2P(\omega_{bd},r) \label{equ:total}
\end{eqnarray}
where we have the definition
\begin{eqnarray}
  &&P(\omega,r)=\overrightarrow{P}(\omega,r)+\overleftarrow{P}(\omega,r) \\
  &&\overrightarrow{P}(\omega,r)=\sum_l G(l)(\big|\overrightarrow{R}(\omega\sqrt{g_{00}}|r)\big|^2) \label{equ:pr}\\
  &&\overleftarrow{P}(\omega,r)=\sum_l G(l)(\big|\overleftarrow{R}(\omega\sqrt{g_{00}}|r)\big|^2)\label{equ:pl}
\end{eqnarray}

According to \eq{equ:total}, we can see that only the term $\omega_b>\omega_d$ has contribution to an atom in the excited, which means that an atom in an excited state can transition to lower level states since the term $\omega_b>\omega_d$ is negative. We can also see that the total rate of change of the atomic energy is zero for an atom in the ground state, which means that the contributions of vacuum fluctuations and radiation reaction cancelled out and the atom in the ground state is stable, which means that spontaneous excitation does not occur in the Boulware vacuum. The \eq{equ:total} is similar to the case that the two-level atom in the flat spacetime and outside the Schwarzschild black hole. The different between them is reflected in the factor $P(\omega_{bd},r)$. Since we cannot solve the exact form of $P(\omega_{bd},r)$, we analyze the behavior of $P(\omega_{bd},r)$ in two asymptotic regions (at the spatial infinity and at the event horizon) to further study the total rate of change of the atomic energy in EGB gravity.

At spatial infinity ($r\rightarrow\infty$), the total rate of change can be obtained by using \eq{equ:23b},\eq{equ:23d} and \eq{equ:total}
\begin{eqnarray}
  \label{equ:inf}
  &&\Big\langle \frac{\ud H_A(\tau)}{\ud \tau} \Big\rangle_{tot}\approx\frac{-e^2}{2\Omega_p}\sum_{\omega_b>\omega_d}|\langle b|\textbf{r}(0)|d\rangle|^2\times \nonumber\\
  &&\Big[\frac{(-2)^{2-p}\pi}{(\Gamma(\frac{p+1}{2}))^2}\cdot \frac{\omega_{bd}^{p+2}p}{p+1} +\sum_l\frac{l(l+p-1)G(l)}{r^{p+2}}|\mathcal{T}_l(\omega_{bd}\sqrt{g_{00}})|^2 \Big]
\end{eqnarray}
From this expression we can see that the second term in the brackets of \eq{equ:inf} tends to be zero at the spatial infinity ($r\rightarrow\infty$). Then only the dimension $p$ has contribution to the total rate of change, and the Boulware vacuum of EGB gravity in high dimension is equivalent to Minkowski vacuum at the spatial infinity.

At the event horizon ($r\rightarrow r_h$), combining \eq{equ:23a}, \eq{equ:23a} and \eq{equ:total}, we can get
\begin{eqnarray}
  &&\Big\langle \frac{\ud H_A(\tau)}{\ud \tau} \Big\rangle_{tot}\approx\frac{-e^2}{2\Omega_p}\sum_{\omega_b>\omega_d}|\langle b|\textbf{r}(0)|d\rangle|^2 \Big[\sum_l\frac{l(l+p-1)G(l)}{ r_h^{p+2}}|\mathcal{T}_l(\omega_{bd}\sqrt{g_{00}})|^2+ \nonumber\\
  &&\frac{\Gamma(\frac{p+2}{2}+\frac{2ir_h\omega_{bd}\sqrt{g_{00}}}{a_1})\Gamma(\frac{p+2}{2}-\frac{2ir_h\omega_{bd}\sqrt{g_{00}}}{a_1})
  \big(\Gamma(\frac{p+2}{2})\big)^2}{\xi^{p+2}r^{p+2}\Gamma(p+2)(p-2)!
  \Gamma(\frac{2ir_h\omega_{bd}\sqrt{g_{00}}}{a_1})\Gamma(-\frac{2ir_h\omega_{bd}\sqrt{g_{00}}}{a_1})}\Big] \label{equ:rh}
\end{eqnarray}
further simplification of the second term in \eq{equ:rh} gives
\begin{subequations}
  \label{equ:sim}
  \begin{numcases}{}
    \prod_{j=0}^m\big[1+(j\cdot\frac{a}{\omega_{bd}})^2\big]\frac{(2r_h\omega_{bd})^{2m+2}\big(\Gamma(m+1)\big)^2}{a_1^{m+1}r^{2m+2}\Gamma(2m+2)(2m-2)!},~p=2m \\
    \prod_{j=0}^m\big[1+(j+\frac{1}{2})^2\cdot\big(\frac{a}{\omega_{bd}}\big)^2\big]\frac{\sqrt{g_{00}}(2r_h\omega_{bd})^{2m+4}\big(\Gamma(\frac{2m+3}{2})\big)^2}
    {a_1^{\frac{2m+5}{2}}r^{2m+3}\Gamma(2m+3)(2m-1)!},~p=2m+1
  \end{numcases}
\end{subequations}
where $a$ represents the proper acceleration of the static atom,
\begin{eqnarray}
  a=\frac{f'(r_h)}{2}\frac{1}{\sqrt{g_{00}}}
\end{eqnarray}

From these results (\eq{equ:rh}, \eq{equ:sim}), we can find that not only the proper acceleration squared term contribute to the total rate of change, but the dimension and the coupling constant $\alpha$ also have contribution to it. We can also see if we let $\alpha\rightarrow 0, p=2$, \eq{equ:rh} will degrade to the result in Ref.\cite{Zhou:2012eb} which calculated a static atom in Schwarzschild spacetime. Since the proper acceleration is divergency ($a\rightarrow\infty$) near the event horizon, the rate of change of the mean energy is also divergency. With some simplification of \eq{equ:sim}, we can get
\begin{eqnarray}
  \label{equ:aa}
  a^p\prod_{j=0}^{[\frac{p-1}{2}]}\big((\frac{\omega_{bd}}{a})^2+(\frac{p}{2}-j)^2\big)
  \frac{2^{p+2}\omega_{bd}^2\big(\Gamma(\frac{p+2}{2})\big)^2}{a_1^{\frac{p}{2}+1}\Gamma(p+2)(p-2)!}
\end{eqnarray}
It is easy to see that $a^p$ is the coefficient of divergency and ($\omega_{bd}/a$) is a dimensionless quantity. Then we can discuss the contribution of the dimension $p$ and the coupling constant $\alpha$ to the total rate of change in three kinds of limit.

\textbf{1.} $\alpha=0$, it will degrade to the Einstein gravity, and
\begin{equation}
  a_1=p-1
\end{equation}
the function \eq{equ:aa} becomes
\begin{eqnarray}
  \label{equ:aa1}
   a^p\prod_{j=0}^{[\frac{p-1}{2}]}\big((\frac{\omega_{bd}}{a})^2+(\frac{p}{2}-j)^2\big)
  \frac{2^{p+2}\omega_{bd}^2\big(\Gamma(\frac{p+2}{2})\big)^2}{(p-1)^{\frac{p}{2}+1}\Gamma(p+2)(p-2)!}
\end{eqnarray}

\textbf{2.} $\alpha\rightarrow\infty$, it will degrade to the pure GB gravity, and
\begin{equation}
  a_1=\frac{1}{2}(p-3)
\end{equation}
the function \eq{equ:aa} becomes
\begin{eqnarray}
  \label{equ:aa2}
   a^p\prod_{j=0}^{[\frac{p-1}{2}]}\big((\frac{\omega_{bd}}{a})^2+(\frac{p}{2}-j)^2\big)
  \frac{2^{p+2}\omega_{bd}^2\big(\Gamma(\frac{p+2}{2})\big)^2}{(\frac{1}{2}(p-3))^{\frac{p}{2}+1}\Gamma(p+2)(p-2)!}
\end{eqnarray}
We can find the coupling constant $\alpha$ has no contribution to the total rate of change in the pure GB gravity.

\textbf{3.}  $\alpha\ll1$, we will get the EGB gravity, and
\begin{eqnarray}
  &&a_1=(p-1)+\frac{\alpha}{r_h^2}(p+1) \\
  &&r_h\approx (2M)^{\frac{1}{p-1}}
\end{eqnarray}
then we can write
\begin{eqnarray}
  \label{equ:b}
  a_1^{\frac{p}{2}+1}=\big((p-1)+\frac{\alpha}{(2M)^{\frac{2}{p-1}}}(p+1)\big)^{\frac{p}{2}+1} \nonumber\\
  \approx (p-1)^{\frac{p}{2}+1}\Big(1+\frac{(p+2)(p+1)}{2(p-1)}\frac{\alpha}{(2M)^{\frac{2}{p-1}}}\Big)
\end{eqnarray}
Comparing with case \textbf{1}($\alpha=0$), there is an additional term relating to the coupling constant in the total rate of change. Having this term, the contribution of the coupling constant to total rate of change is clear.

\subsection{Unruh vacuum}

Substituting \eq{equ:ee1} to \eq{equ:cf} and \eq{equ:xf}, the symmetric correlation function and the linear susceptibility function can be written as
\begin{eqnarray}
  C^F(x(\tau),x(\tau'))&&=\frac{1}{8\pi\Omega_p}\int_0^{\infty}\ud\omega ~\omega(e^{-\frac{i\omega\Delta\tau}{\sqrt{g_{00}}}}+e^{\frac{i\omega\Delta\tau}{\sqrt{g_{00}}}}) \nonumber\\
  &&\times\Big[\frac{\sum_l G(l)|\overrightarrow{R}|^2}{1-e^{-2\pi\omega/\kappa}}
  +\theta(\omega)\sum_l G(l)|\overleftarrow{R}|^2 \Big] \\
  \chi^F(x(\tau),x(\tau'))&&=\frac{1}{8\pi\Omega_p}\int_0^{\infty}\ud\omega ~\omega(e^{-\frac{i\omega\Delta\tau}{\sqrt{g_{00}}}}-e^{\frac{i\omega\Delta\tau}{\sqrt{g_{00}}}}) \nonumber\\
  &&\times\Big[\frac{\sum_l G(l)|\overrightarrow{R}|^2}{1-e^{-2\pi\omega/\kappa}}
  +\theta(\omega)\sum_l G(l)|\overleftarrow{R}|^2 \Big]
\end{eqnarray}

Using the same procedure mentioned above, the total rate of change of the mean atomic energy can be written as
\begin{eqnarray}
  \Big\langle\frac{\ud H_A(\tau)}{\ud \tau}\Big\rangle_{tot}&&=-\frac{e^2 g_{00}}{2\Omega_p}\Big\{\sum_{\omega_b>\omega_d}|\langle b|r(0)|d\rangle|^2\omega_{bd}^2 \Big[\Big(1+\frac{1}{e^{2\pi\omega_{bd}/\kappa_r}-1} \Big)\overrightarrow{P}(\omega_{bd},r)+\overleftarrow{P}(\omega_{bd},r) \Big] \nonumber\\
  &&-\sum_{\omega_b<\omega_d}|\langle b|r(0)|d\rangle|^2\omega_{bd}^2\frac{\overrightarrow{P}(\omega_{bd},r)}{e^{2\pi|\omega_{bd}|/\kappa_r}-1} \Big\} \label{equ:total1}
\end{eqnarray}
where $\kappa_r=\kappa/\sqrt{g_{00}}$ and the symbols $\overrightarrow{P}$ and $\overleftarrow{P}$ have the same definition with \eq{equ:pr} and \eq{equ:pl}.

Comparing with the total rate of change in Boulware vacuum \eq{equ:total}, the term $\omega_b<\omega_d$ gives a positive contribution to the total rate of change, which means the atom in the ground state will spontaneous excite. For the same reason, we study the behavior of $P(\omega_{bd},r)$ in two asymptotic regions  ($r\rightarrow\infty$ and $r\rightarrow r_h$).

When the atom is placed at spatial infinity ($r\rightarrow\infty$), the total rate of change can be obtained by using \eq{equ:23b},\eq{equ:23d} and \eq{equ:total1}
\begin{eqnarray}
  \label{equ:inf1}
  &&\Big\langle \frac{\ud H_A(\tau)}{\ud \tau} \Big\rangle_{tot}\approx\frac{-e^2}{2\Omega_p}\Big\{\sum_{\omega_b>\omega_d}|\langle b|\textbf{r}(0)|d\rangle|^2 \Big[\Big(1+\frac{1}{e^{2\pi\omega_{bd}/\kappa_r}-1} \Big) \nonumber\\
  &&\times\sum_l\frac{l(l+p-1)G(l)}{r^{p+2}}|\mathcal{T}_l(\omega_{bd}\sqrt{g_{00}})|^2
  +\frac{(-2)^{2-p}\pi}{(\Gamma(\frac{p+1}{2}))^2}\cdot \frac{\omega_{bd}^{p+2}p}{p+1} \Big] \nonumber\\
  &&-\sum_{\omega_b<\omega_d}\frac{|\langle b|r(0)|d\rangle|^2}{e^{2\pi|\omega_{bd}|/\kappa_r}-1}\sum_l\frac{l(l+p-1)G(l)}{r^{p+2}}|\mathcal{T}_l(\omega_{bd}\sqrt{g_{00}})|^2 \Big\}
\end{eqnarray}
Since the reflection coefficient $\mathcal{T}$ tends to be zero at spatial infinity, we can see that the degeneration of \eq{equ:inf1} is similar with the case of Boulware vacuum (\eq{equ:inf}), only the dimension $p$ has contribution to the total rate of change, which shows that the Unruh vacuum of EGB gravity in high dimension is equivalent to Minkowski vacuum at the spatial infinity as well.

When the atom is fixed near the event horizon ($r\rightarrow r_h$), combining \eq{equ:23a}, \eq{equ:23a} and \eq{equ:total1}, we can get
\begin{eqnarray}
  \label{equ:inf11}
  &&\Big\langle \frac{\ud H_A(\tau)}{\ud \tau} \Big\rangle_{tot}\approx\frac{-e^2}{2\Omega_p}\Big\{\sum_{\omega_b>\omega_d}|\langle b|\textbf{r}(0)|d\rangle|^2 \Big[\Big(1+\frac{1}{e^{2\pi\omega_{bd}/\kappa_r}-1} \Big) \nonumber\\
  &&\times\Big(a^p\prod_{j=0}^{[\frac{p-1}{2}]}\big((\frac{\omega_{bd}}{a})^2+(\frac{p}{2}-j)^2\big)
  \frac{2^{p+2}\omega_{bd}^2\big(\Gamma(\frac{p+2}{2})\big)^2}{a_1^{\frac{p}{2}+1}\Gamma(p+2)(p-2)!}\Big)
  +\sum_l\frac{l(l+p-1)G(l)}{ r_h^{p+2}}|\mathcal{T}_l(\omega_{bd}\sqrt{g_{00}})|^2 \Big] \nonumber\\
  &&-\sum_{\omega_b<\omega_d}\frac{|\langle b|r(0)|d\rangle|^2}{e^{2\pi|\omega_{bd}|/\kappa_r}-1}\Big(a^p\prod_{j=0}^{[\frac{p-1}{2}]}\big((\frac{\omega_{bd}}{a})^2+(\frac{p}{2}-j)^2\big)
  \frac{2^{p+2}\omega_{bd}^2\big(\Gamma(\frac{p+2}{2})\big)^2}{a_1^{\frac{p}{2}+1}\Gamma(p+2)(p-2)!} \Big) \Big\}
\end{eqnarray}
We can find it is the same as the case of Boulware vacuum that the dimension, the coupling constant and the proper acceleration have contribution to the total rate of change. While the different is that the Hawking radiation can be found in Unruh vacuum and the corresponding temperature $T=\kappa_r/2\pi$, which leads to a remarkable observation that for an atom in the ground state, the spontaneous excitation can appear due to the Unruh effect. The influence of the parameters of the dimension and the coupling constant to the spontaneous excitation is the same as the discussion above (\eq{equ:aa} to \eq{equ:b}).

\subsection{Hartle-Hawking vacuum}

Substituting \eq{equ:ee2} to \eq{equ:cf} and \eq{equ:xf}, we can obtain the symmetric correlation function and the linear susceptibility function in Hartle-Hawking vacuum
\begin{eqnarray}
  C^F(x(\tau),x(\tau'))&&=\frac{1}{8\pi\Omega_p}\int_0^{\infty}\ud\omega ~\omega(e^{-\frac{i\omega\Delta\tau}{\sqrt{g_{00}}}}+e^{\frac{i\omega\Delta\tau}{\sqrt{g_{00}}}}) \nonumber\\
  &&\times\Big[\frac{\sum_l G(l)|\overrightarrow{R}|^2}{1-e^{-2\pi\omega/\kappa}}
  +\frac{\sum_l G(l)|\overleftarrow{R}|^2}{e^{2\pi\omega/\kappa}-1} \Big] \\
  \chi^F(x(\tau),x(\tau'))&&=\frac{1}{8\pi\Omega_p}\int_0^{\infty}\ud\omega ~\omega(e^{-\frac{i\omega\Delta\tau}{\sqrt{g_{00}}}}-e^{\frac{i\omega\Delta\tau}{\sqrt{g_{00}}}}) \nonumber\\
  &&\times\Big[\frac{\sum_l G(l)|\overrightarrow{R}|^2}{1-e^{-2\pi\omega/\kappa}}
  -\frac{\sum_l G(l)|\overleftarrow{R}|^2}{e^{2\pi\omega/\kappa}-1} \Big]
\end{eqnarray}

Using the same procedure of Boulware vacuum, we can write the total rate of change of the mean atomic energy as
\begin{eqnarray}
  \Big\langle\frac{\ud H_A(\tau)}{\ud \tau}\Big\rangle_{tot}&&=-\frac{e^2 g_{00}}{2\Omega_p}\Big\{\sum_{\omega_b>\omega_d}|\langle b|r(0)|d\rangle|^2\omega_{bd}^2 \Big(1+\frac{1}{e^{2\pi\omega_{bd}/\kappa_r}-1} \Big) \nonumber\\
  &&\times\big(\overrightarrow{P}(\omega_{bd},r)+\overleftarrow{P}(-\omega_{bd},r) \big)-\sum_{\omega_b<\omega_d}|\langle b|r(0)|d\rangle|^2\omega_{bd}^2 \nonumber\\
  &&\times \frac{1}{e^{2\pi|\omega_{bd}|/\kappa_r}-1}\big(\overrightarrow{P}(\omega_{bd},r)+\overleftarrow{P}(-\omega_{bd},r)\big) \Big\} \label{equ:total2}
\end{eqnarray}
We can find the positive term $\omega_b<\omega_d$ also appears in the Hartle-Hawking vacuum, which will lead to spontaneous excitation of the atom in the ground state.

When the atom is placed at spatial infinity ($r\rightarrow\infty$), the total rate of change can be obtained by using \eq{equ:23b},\eq{equ:23d} and \eq{equ:total2}
\begin{eqnarray}
  \label{equ:inf2}
  \Big\langle\frac{\ud H_A(\tau)}{\ud \tau}\Big\rangle_{tot}&&\approx\frac{-e^2 }{2\Omega_p}\Bigg\{\sum_{\omega_b>\omega_d}|\langle b|r(0)|d\rangle|^2 \Big(1+\frac{1}{e^{2\pi\omega_{bd}/\kappa_r}-1} \Big)
  \Big[\frac{2^{2-p}\pi}{(\Gamma(\frac{p+1}{2}))^2}\cdot \frac{\omega_{bd}^{p+2}p}{p+1}  \nonumber\\
  &&+\sum_l\frac{l(l+p-1)G(l)}{r^{p+2}}|\mathcal{T}_l(\omega_{bd}\sqrt{g_{00}})|^2 \Big]-\sum_{\omega_b<\omega_d} \frac{|\langle b|r(0)|d\rangle|^2}{e^{2\pi|\omega_{bd}|/\kappa_r}-1}  \nonumber\\
  &&\times\Big[\frac{2^{2-p}\pi}{(\Gamma(\frac{p+1}{2}))^2}\cdot \frac{\omega_{bd}^{p+2}p}{p+1} +\sum_l\frac{l(l+p-1)G(l)}{r^{p+2}}|\mathcal{T}_l(\omega_{bd}\sqrt{g_{00}})|^2 \Big] \Bigg\}
\end{eqnarray}
From \eq{equ:inf2}, it shows that the Hartle-Hawking vacuum is equivalent to the Minkowski vacuum with the Hawking temperature $T=\kappa_r/2\pi$ at the spatial infinity, which means that the atom which is placed at spatial infinity in Hartle-Hawking vacuum is equivalent to be immersed in a thermal bath at temperature $T$ in Minkowski vacuum. We can also find that in this region the spontaneous excitation occur for a static atom in the ground state.

For the atom which is fixed near the event horizon $(r\rightarrow r_h)$, combining \eq{equ:23a}, \eq{equ:23a} and \eq{equ:total2}, we can obtain
\begin{eqnarray}
  \label{equ:inf22}
  &&\Big\langle\frac{\ud H_A(\tau)}{\ud \tau}\Big\rangle_{tot}\approx\frac{-e^2 }{2\Omega_p}\Bigg\{\sum_{\omega_b>\omega_d}|\langle b|r(0)|d\rangle|^2 \Big(1+\frac{1}{e^{2\pi\omega_{bd}/\kappa_r}-1} \Big) \nonumber\\
  &&\times\Big[\sum_l\frac{l(l+p-1)G(l)}{r^{p+2}}|\mathcal{T}_l(\omega_{bd}\sqrt{g_{00}})|^2
  +a^p\prod_{j=0}^{[\frac{p-1}{2}]}\big((\frac{\omega_{bd}}{a})^2+(\frac{p}{2}-j)^2\big)
  \frac{2^{p+2}\omega_{bd}^2\big(\Gamma(\frac{p+2}{2})\big)^2}{a_1^{\frac{p}{2}+1}\Gamma(p+2)(p-2)!} \Big] \nonumber\\
  &&-\sum_{\omega_b<\omega_d} \frac{|\langle b|r(0)|d\rangle|^2}{e^{2\pi|\omega_{bd}|/\kappa_r}-1}
  \Big[\sum_l\frac{l(l+p-1)G(l)}{r^{p+2}}|\mathcal{T}_l(\omega_{bd}\sqrt{g_{00}})|^2  \nonumber\\
  &&+a^p\prod_{j=0}^{[\frac{p-1}{2}]}\big((\frac{\omega_{bd}}{a})^2+(\frac{p}{2}-j)^2\big)
  \frac{2^{p+2}\omega_{bd}^2\big(\Gamma(\frac{p+2}{2})\big)^2}{a_1^{\frac{p}{2}+1}\Gamma(p+2)(p-2)!} \Big] \Bigg\}
\end{eqnarray}
Comparing \eq{equ:inf22} with \eq{equ:inf11}, we can find that there is an additional reflection term which represents the incoming thermal radiation from infinity. Combining \eq{equ:inf2} and \eq{equ:inf22}, it is easy to see that the Hartle-Hawking vacuum is not empty at infinity and the spontaneous excitation can occur near the event horizon either.

\section{CONCLUSION}

In this paper, using the Gupta-Bleuler quantization of free electromagnetic fields in a static spherically symmetric spacetime of arbitrary dimension, we calculated the two-point functions for the electromagnetic fields in Boulware vacuum of $d$-dimensional EGB gravity and analyzed their properties in two asymptotic regions. We also gave the contributions of vacuum fluctuations and radiation reaction to the total rate of change of the mean energy for a two-level atom in interaction with quantum electromagnetic fluctuations in Boulware vacuum, Unruh vacuum and Hartle-Hawking vacuum. We found that spontaneous excitation does not occur in Boulware vacuum but occurred in the Unruh vacuum and Hartle-Hawking vacuum.

The expression of the total rate of change of the mean energy shows that in Boulware vacuum it is stable for a static atom in the ground state and it will transit to lower level states for the static atom in an excited state, in Unruh vacuum and Hartle-Hawking vacuum the balance between the contributions of vf and rr no longer exists for the atom in the ground state and the spontaneous excitation occurs. This conclusion is in qualitative agreement with the case that the atom interacts with quantized massless scalar fields (\cite{Yu:2007wv}) or quantized electromagnetic fields outside a four-dimensional Schwarzschild black hole. However, there are some differences between them. In four-dimensional Schwarzschild black hole, the spontaneous emission rate of the excited atoms near the horizon contains an extra term proportional to the squared proper acceleration of the atom. In $d$-dimensional EGB black hole, it also contains a similar extra term, but proportional to a polynomial by coupling the squared proper acceleration and the dimension of the spacetime. Finally, we respectively discuss the contribution of the coupling constant and the dimension to the total rate of change when the coupling constant takes three different limit which represent three kinds of different spacetime.

\section{Acknowledgments}
This work has been supported by the National Natural Science Foundation of China (NSFC)
under Grant Nos. 11275099 and 11347605.

\end{document}